# Current-in-plane spin-valve magnetoresistance in ferromagnetic semiconductor (Ga,Fe)Sb heterostructures with high Curie temperature

Kengo Takase,[1] Le Duc Anh,[1,2] Kosuke Takiguchi,[1] and Masaaki Tanaka[1,3,*]

[1]*Dept. of Electrical Engineering and Information Systems, The University of Tokyo,*
*7-3-1 Hongo, Bunkyo-ku, Tokyo 113-8656, Japan*
[2]*Institute of Engineering Innovation, The University of Tokyo,*
*7-3-1 Hongo, Bunkyo-ku, Tokyo 113-8656, Japan*
[3]*Center for Spintronics Research Network (CSRN), The University of Tokyo,*
*7-3-1 Hongo, Bunkyo-ku, Tokyo 113-8656, Japan*
*E-mail: masaaki@ee.t.u-tokyo.ac.jp

**Abstract**

We demonstrate spin-valve magnetoresistance with a current-in-plane (CIP) configuration in (Ga,Fe)Sb / InAs (thickness $t_{InAs}$ nm) / (Ga,Fe)Sb trilayer heterostructures, where (Ga,Fe)Sb is a ferromagnetic semiconductor (FMS) with high Curie temperature ($T_C$). An MR curve with an open minor loop is clearly observed at 3.7 K in a sample with $t_{InAs}$ = 3 nm, which originates from the parallel - antiparallel magnetization switching of the (Ga,Fe)Sb layers and spin-dependent scattering at the (Ga,Fe)Sb / InAs interfaces. The MR ratio increases (from 0.03 to 1.6%) with decreasing $t_{InAs}$ (from 9 to 3 nm) due to the enhancement of the interface scattering. This is the first demonstration of the spin-valve effect in Fe-doped FMS heterostructures, paving the way for device applications of these high- $T_C$ FMSs.

Spin-valve structures, consisting of ferromagnetic/nonmagnetic multilayers, have attracted great interest as fundamental building blocks in spintronic devices owing to their magnetoresistances (MRs), such as giant magnetoresistance (GMR) and tunneling magnetoresistance (TMR).[1–4] Such MRs have been widely studied using ferromagnetic metals, ferromagnetic nanoparticles,[5,6] and ferromagnetic semiconductors (FMSs).[7–11] Comparing with spin-valve structures based on ferromagnetic metals and nanoparticles, those based on FMSs are fully epitaxially grown on semiconductor substrates and possess excellent crystal quality and compatibility with the existing semiconductor technology. Thus, we can immensely benefit from the well-established knowledge and techniques of semiconductor materials and devices, such as band engineering. Another important feature is their high controllability of carrier-induced properties: Using a gate voltage, one can control either the electrical transport in the nonmagnetic semiconductor channel or the magnetic properties of the FMS layers[12-16]. These advantages can be utilized to realize non-volatile spintronic devices such as spin transistors[17,18] and spin diodes,[19,20,21] which are essential for low-power electronics.

Some pioneering studies on the spin-valve structures based on FMSs were conducted using Mn-doped FMSs such as (Ga,Mn)As and (In,Mn)As, which are well-known FMSs. In addition to the observations of GMR[7] and TMR,[8,9] vertical spin MOSFET operation was demonstrated using (Ga,Mn)As.[11] Nevertheless, these Mn-doped FMSs are ferromagnetic only at very low temperature (the maximum Curie temperature ($T_C$) of Mn-doped FMSs is only 200 K[22]). Furthermore, only p-type FMSs are available as long as one chooses Mn as the magnetic dopant because Mn atoms are acceptors in III-V semiconductors. These shortcomings seriously limit their potential applications in future low-power-consumption and room-temperature spin devices.

In this work, we investigate spin-valve structures of a new class of FMSs, in which we use Fe as the magnetic dopant in III-V semiconductors, namely Fe-doped III-V FMSs. These Fe-doped III-V FMSs are both p-type [(Ga,Fe)Sb, (Al,Fe)Sb][23,24] and n-type [(In,Fe)As, (In,Fe)Sb][25,26], some of which are found to have $T_C$ higher than room temperature and thus overcome the limitations of the Mn-doped FMSs. In addition, a large spontaneous spin-splitting in their band structures and strong *s(p)-d* exchange interactions provide notable benefits for device applications by utilizing band and wavefunction engineering.[20,27,28] These unique features make the Fe-based FMSs one of the most promising material platforms for future spin-based electronics. Towards realization of practical spintronic devices of these Fe-doped FMSs, demonstration of the spin-valve effects such as GMR and TMR using these materials will be an important milestone; however, it is still lacking.

To realize and study the spin-valve effect in the Fe-doped FMSs, we choose (Ga,Fe)Sb / InAs / (Ga,Fe)Sb trilayer heterostructures. (Ga,Fe)Sb is a p-type FMS with high $T_C$,[29,30] while InAs is a nonmagnetic channel with high electron mobility. The lattice mismatch between InAs and (Ga,Fe)Sb is only ~0.1%[29], which enables an epitaxial growth of high-quality heterostructures. Furthermore, InAs / (Ga,Fe)Sb is a type-III heterostructure, i.e., the conduction band bottom of InAs is lower than the valence band top of (Ga,Fe)Sb at the interface. This staggered band profile enables large penetration of the electron wavefunction in InAs into the (Ga,Fe)Sb side, and consequently a strong interfacial magnetic proximity effect (MPE).[31] The detailed structure consists of (from top to bottom) $(Ga_{0.75},Fe_{0.25})Sb$ (40 nm) / InAs ($t_{InAs}$ nm)/$(Ga_{0.8},Fe_{0.2})Sb$ (40 nm) / AlSb (150 nm) / AlAs (5 nm) / GaAs buffer (150 nm) on a semi-insulating GaAs (001) substrate grown by molecular beam epitaxy (MBE) (see Supplementary Material). The (Ga,Fe)Sb layers are designed to have a high $T_C$ (= 260 K) and a perpendicular magnetic anisotropy at low temperature.[29,30,32] We vary the thickness $t_{InAs}$ of InAs to be 0, 3, 6, 9 nm (denoted as sample A, B, C, D, respectively). Sample A, which does not contain an InAs layer, serves as a reference. *In situ* reflection high energy electron diffraction (RHEED) patterns are streaky, as shown in Fig. 1(a) – (c), indicating a two-dimensional growth mode and zinc-blende structure maintained throughout the MBE growth. The samples are patterned into Hall bar devices with size of $100 \times 400$ μm² using standard photolithography and Ar ion milling. Magneto-transport measurements are performed by a four-point method, with a current flown in the film plane and a magnetic field ***H*** applied perpendicular to the film plane [Fig. 1(e)].

In the present (Ga,Fe)Sb / InAs / (Ga,Fe)Sb heterostructures, the Fermi level ($E_F$) is pinned in the band-gap of (Ga,Fe)Sb layers due to the formation of the Fe-related impurity band.[28] Therefore electron carriers are accumulated and confined in the InAs layer due to the potential barrier (0.5 ~ 0.6 eV) from the conduction band bottom of (Ga,Fe)Sb [Fig. 1(d)]. Longitudinal resistances $R$ of all the samples show monotonic increase with decreasing temperature, as shown in Fig. 2(a). Comparing with the reference sample A where there is no InAs, $R$ drops significantly and the samples show more metallic-like behavior when we increase the InAs thickness. At low temperature (< 20 K), the $R$ values of sample B ($t_{InAs}$ = 3 nm) and sample C, D ($t_{InAs}$ = 6, 9 nm) are two and three orders of magnitude smaller than that of the reference sample A ($t_{InAs}$ = 0 nm), respectively. These results indicate that the InAs layer is responsible for over 99% and the (Ga,Fe)Sb layers only contribute less than 1% of the electrical transport in sample B, C, and D at low temperature.

Hall resistances of all the samples are shown in Fig. 2(b) – (e). Sample A, which has no InAs channel, shows a strong anomalous Hall effect (AHE) originated from the (Ga,Fe)Sb layers [Fig. 2(b)]. Interestingly, sample B ($t_{InAs}$ = 3 nm), where more than 99% of electrical transport occurs in the InAs channel, also shows an AHE whose value is the same order as that of sample A [Fig. 2(c)]. In this sample, the MPE from the two (Ga,Fe)Sb layers causes a spontaneous spin-splitting and strong interfacial spin-dependent scattering in the InAs layer, as reported in (Ga,Fe)Sb / InAs bilayers.[32] This MPE may be the origin of the observed AHE in sample B. On the other hand, in sample C and D, linear negative Hall resistances indicate that the magnetic proximity effect decays quickly with increasing $t_{InAs}$, and n-type conduction in the InAs layer becomes completely dominant [Fig. 2(d)(e)]. The electron carrier concentrations $n$ estimated from the Hall resistances are $n$ = 3.2×10$^{12}$ cm$^{-2}$ for sample C and 4.4×10$^{12}$ cm$^{-2}$ for sample D at 3.7 K, which are about five times larger than that of an undoped GaSb / InAs / GaSb QW.[33] This is possibly due to the higher density of defects formed in the trilayers because of the low growth temperature (see Supplementary Material), and the higher pinning position of the Fermi level in InAs due to the formation of the Fe-related impurity band in the (Ga,Fe)Sb layers as shown in Fig. 1(d).

Fig. 3(a) shows MR curves at 3.7 K of all the samples when a magnetic field ***H*** is applied in the range of (–10, 10 kOe). The MR ratio is defined as $(R - R_{min})/R_{min}$, where $R_{min}$ is the minimum resistance value. A spin-valve-like MR, which is dominant at low magnetic field, is superimposed with a background MR, which almost linearly depends on the magnetic field and has a positive or negative slope depending on $t_{InAs}$. The negative background MR in the sample C and D ($t_{InAs}$ = 6 and 9 nm), where the current dominantly flows in the InAs channel, is attributed to the MPE reported in (Ga,Fe)Sb / InAs bilayer.[31] This negative MR can be fitted and excluded using the modified Khosla-Fischer model. On the other hand, the positive linear background MR in sample A is a property of the (Ga$_{0.75}$,Fe$_{0.25}$)Sb / (Ga$_{0.8}$,Fe$_{0.2}$)Sb bilayer, whose origin is unknown. We note that a similar giant linear positive MR was reported in a magnetic nanocolumnar GeMn system.[34] In the case of heavily doped (Ga,Fe)Sb, columnar Fe-rich (Ga,Fe)Sb regions are formed due to spinodal decomposition.[31] This suggest that the positive MR may originate from the same mechanism. For the sample B ($t_{InAs}$ = 3 nm) only the spin-valve-like MR is observed.

To estimate the spin-valve MR signal, we define the spin-valve MR ratio $MR_{SV} \equiv (R_{max} - R_{min})/R_{min}$, where $R_{max}$ ($R_{min}$) represents a maximum (minimum) value of $R$ after excluding the background MR (see Supplementary Materials). The $MR_{SV}$ due to the spin-valve effect becomes larger (from 0.03 to 1.6%) with decreasing $t_{InAs}$ (from 9 to 3 nm) as shown in Fig. 3(b). The increase of $MR_{SV}$ clearly reflects the enhancement of the spin-dependent interface scattering with decreasing the thickness of the conducting

InAs layer. $MR_{SV}$ of the sample B reaches 1.6%, which is an order of magnitude larger than the previous studies of CIP-GMR based on (Ga,Mn)As.[7,35,36] It should be noted that there is also a sizable spin-valve-like MR (~2%) in the sample A ($t_{InAs}$ = 0 nm), which originates either from an anisotropic MR (AMR) of the (Ga,Fe)Sb[29] or a GMR in granular systems.[5,6] In sample B, C, and D, in which the (Ga,Fe)Sb layers contribute less than 1% in the total electrical resistance, the MR contribution of (Ga,Fe)Sb to the $MR_{SV}$ is less than 0.02% and should be negligible.

We show in the upper panel of Fig. 3(c) detailed results of major loop and minor loop measurements of the spin-valve $MR_{SV}$ in sample B ($t_{InAs}$ = 3 nm) at 3.7 K. We also show in the lower panel of Fig. 3(c) the magnetization hysteresis of the trilayer measured by AHE and superconducting quantum interference device (SQUID) magnetometry. It is difficult to distinguish the difference of the coercive forces of the top and bottom (Ga,Fe)Sb layers from the SQUID result. We note that the SQUID result was obtained at a slightly higher temperature (5 K), which possibly leads to a small deviation from the magnetic hysteresis measured by AHE (measured at 3.7 K). The coercive forces of AHE coincide with the peaks of the MR, which supports our conclusion that the $MR_{SV}$ results from the parallel-antiparallel (P-AP) magnetization switching of the (Ga,Fe)Sb layers. The clear and open minor loop curve (green curve) indicates that the P and AP magnetization configurations of the top and bottom (Ga,Fe)Sb can be stably established. The realization of resistance difference in the P and AP configurations is an important milestone for non-volatile spin device applications of (Ga,Fe)Sb.

MR curves of sample B ($t_{InAs}$ = 3 nm) measured at higher temperatures are shown in Fig. 4. We also show the magnetic hysteresis curves measured by magnetic circular dichroism (MCD), which mostly reflects the magnetization of the top (Ga,Fe)Sb layer. As we increase temperature, the remanent magnetization of the top (Ga,Fe)Sb decreases quickly to almost zero above 100 K. This is because the perpendicular magnetic anisotropy (PMA) of (Ga,Fe)Sb weakens with increasing temperature.[37] Simultaneously, the MR ratio becomes small and almost vanishes above 200 K. To obtain a sizable spin-valve effect at room temperature, (Ga,Fe)Sb thin films with larger remanent magnetization at high temperature are definitely required. The spin-valve effect can also be enhanced in a current-perpendicular-to-plane (CPP) configuration, as reported in other magnetic material systems.[38]

In summary, we have demonstrated a clear MR (~1.6%) due to the spin-valve effect in trilayer heterostructures containing high-$T_C$ FMS (Ga,Fe)Sb. From the major loop and minor loop MRs, we concluded the origin of the observed MRs is the spin-valve effect that originates from the spin-dependent scattering at the (Ga,Fe)Sb/InAs interfaces.

The MR ratio increases (from 0.03 to 1.6%) with decreasing $t_{\text{InAs}}$ (from 9 to 3 nm). The demonstration of the spin-valve effect in Fe-doped FMSs in this work is the first important step towards device applications of these high-$T_\text{C}$ FMSs.

**Acknowledgements**
This work was supported by Grants-in-Aid for Scientific Research (Grant Nos. 16H02095, 17H04922, 18H05345, and 19K21961), CREST (No. JPMJCR1777) and PRESTO Programs (No. JPMJPR19LB) of JST. Part of this work was carried out under the support of Spintronics Research Network of Japan (Spin-RNJ).

**Data Availability**
The data that support the findings of this study are available from the corresponding author upon reasonable request.

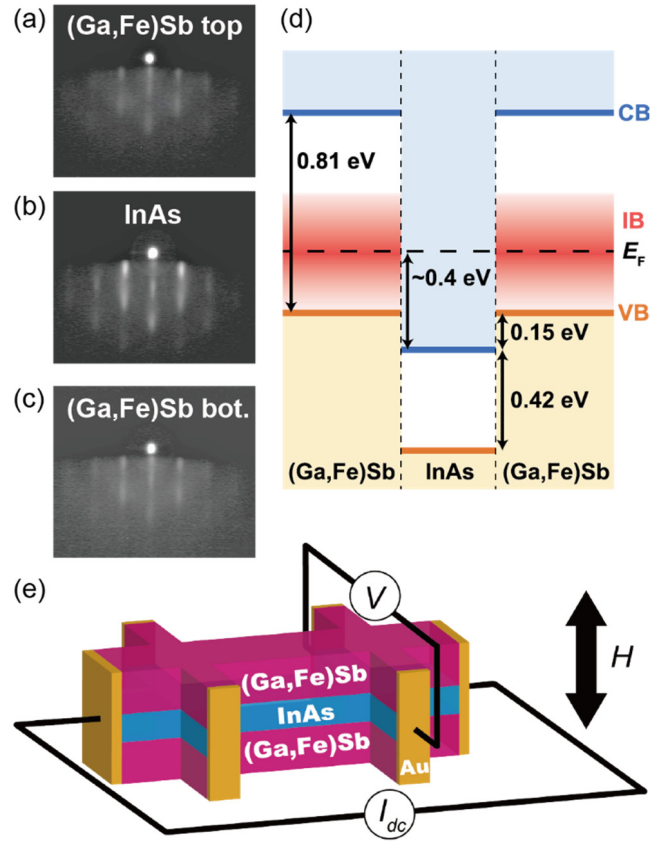

**Fig. 1**. (a) – (c) Reflection high-energy electron diffraction (RHEED) patterns taken along the $[\bar{1}10]$ azimuth during the MBE growth of sample C. (d) Schematic band alignment of the (Ga,Fe)Sb / InAs / (Ga,Fe)Sb heterostructures. CB, IB, VB denote the conduction band, Fe-related impurity band, and valence band. The impurity band lies in the bandgap of the heavily Fe-doped (Ga,Fe)Sb, as shown by the gradient-red color. The Fermi level $E_F$ is pinned at 0.5 – 0.6 eV below the conduction band bottom of the (Ga,Fe)Sb.[28] (e) Schematic device structure and measurement configuration of the Hall bar devices with size of $100 \times 400$ μm$^2$.

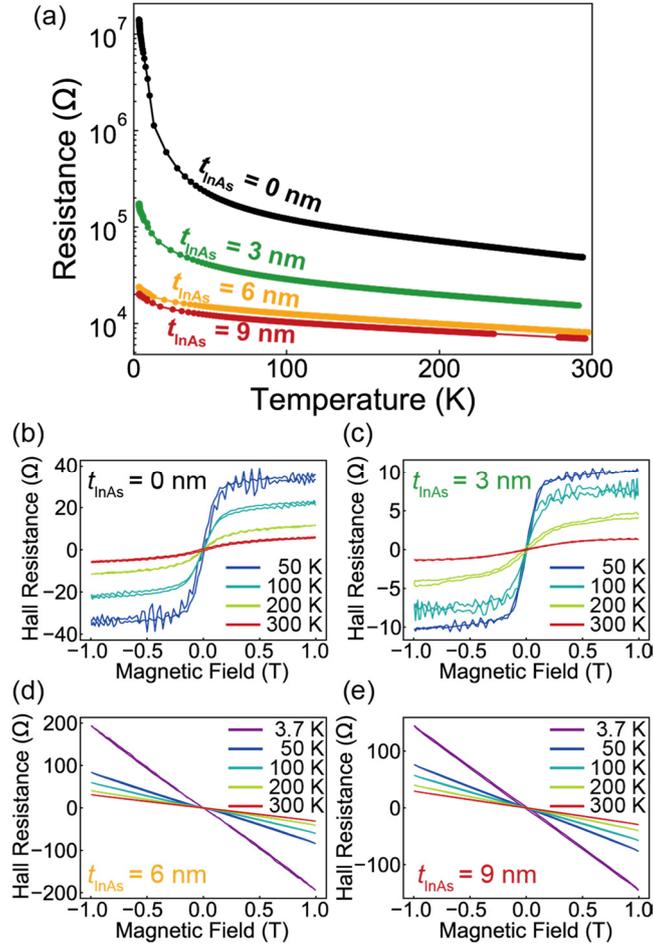

**Fig. 2**. (a) Temperature dependence of the longitudinal resistance at zero magnetic field, and (b) – (e) temperature dependence of Hall resistances in all the samples. The electron carrier concentrations $n$ estimated from the Hall resistance are $n = 3.2 \times 10^{12}$ cm$^{-2}$ for sample C and $n = 4.4 \times 10^{12}$ cm$^{-2}$ for sample D at 3.7 K. The carrier concentrations of sample A and B cannot be estimated due to the strong AHE.

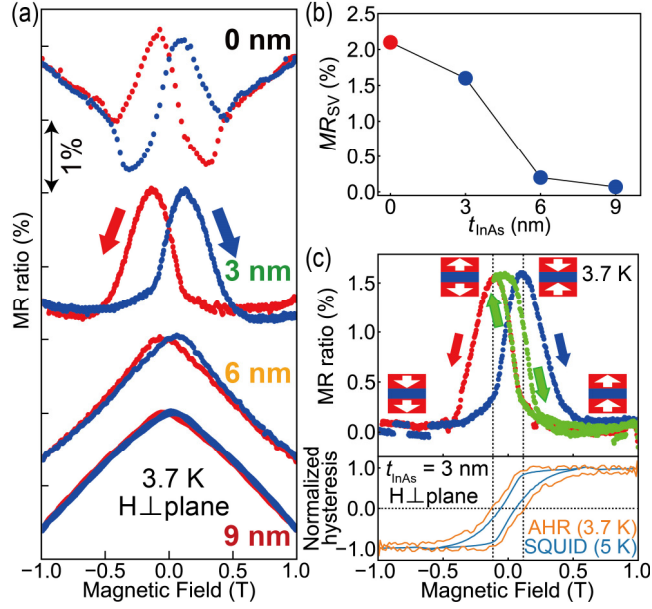

**Fig. 3**. (a) MR curves of the samples A – D measured at 3.7 K with a magnetic field ***H*** applied perpendicular to the film plane. The red and blue curves are the major loops with magnetic-field sweeping directions of + to – and – to +, respectively. (b) Dependence of $MR_{SV}$ on $t_{InAs}$. Blue and red circles represent the MR originates from the spin-valve effect and GMR in a granular system, respectively. (c) Detailed results of major loop and minor loop (green curve) measurements of the spin-valve $MR_{SV}$ (upper panel) and the corresponding magnetization hysteresis curves (lower panel) in sample B ($t_{InAs}$ = 3 nm), measured by anomalous Hall resistance (AHR) at 3.7 K and SQUID at 5 K. White arrows show parallel and anti-parallel magnetization configurations.

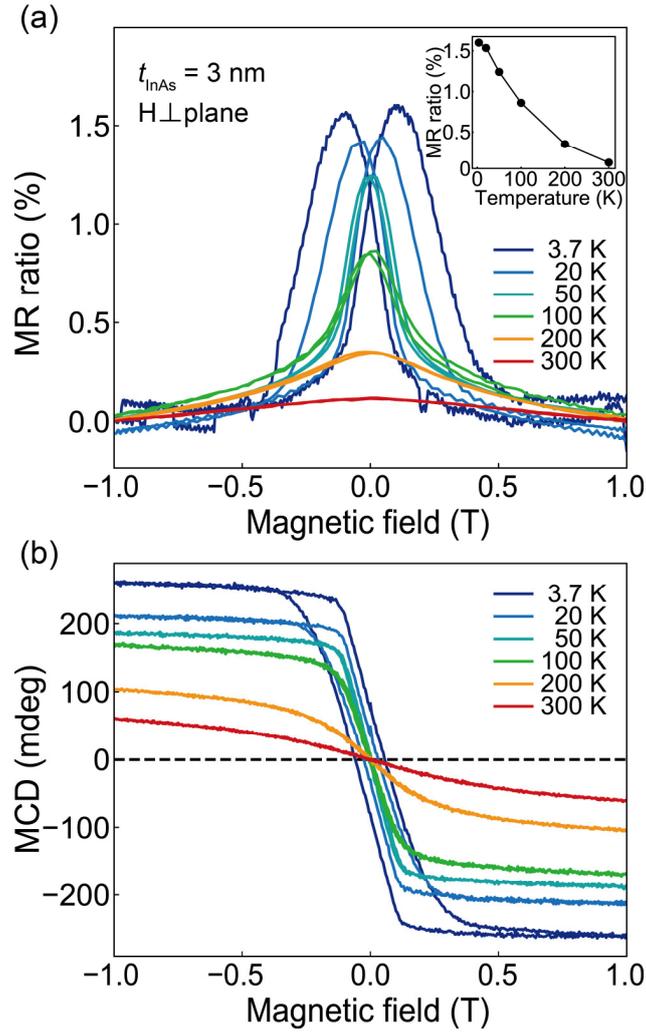

**Fig. 4**. (a) MR curves of sample B ($t_{InAs}$ = 3 nm) measured at various temperatures. Inset is the MR ratio vs. temperature. (b) Magnetization hysteresis curves measured by magnetic circular dichroism (MCD), which reflects the magnetization of the top (Ga,Fe)Sb layer of sample B at various temperatures.

# Supplementary material

**Current-in-plane spin-valve magnetoresistance effect in ferromagnetic semiconductor (Ga,Fe)Sb heterostructures with high Curie temperature**

1. **Sample growth**

We grew heterostructures consisting of (from top to bottom) $(Ga_{0.75},Fe_{0.25})Sb$ (40 nm) / InAs ($t_{InAs}$ nm) / $(Ga_{0.8},Fe_{0.2})Sb$ (40 nm) / AlSb (150 nm) / AlAs (5 nm) / GaAs buffer (150 nm) on a semi-insulating GaAs (001) substrate grown by molecular beam epitaxy (MBE) [Fig. S1]. The growth temperature was 550°C for the GaAs and AlAs layers, 470°C for the AlSb layer, and 250°C for the top (Ga,Fe)Sb / InAs / (Ga,Fe)Sb trilayers. The growth rate in all the layers is 500 nm/h. Fluxes of the group III elements (Ga, Al, In) were calibrated by monitoring oscillations of the reflection high energy electron diffraction (RHEED) intensity during the MBE growth. The Fe flux was calibrated using secondary ion mass spectrometry (SIMS) combined with Rutherford back scattering (RBS) in a reference sample. The insulating AlSb layer is used to relax the large lattice mismatch (~7%) between the top trilayers and the GaAs substrate. Different Fe concentrations (25% vs. 20%) in the top and bottom (Ga,Fe)Sb layers were chosen to yield different coercive forces in the magnetization curves. At these Fe concentrations, a thickness of 40 nm is necessary to obtain a perpendicular magnetization axis in these two FMS layers at low temperature (ref. 32). *In situ* RHEED patterns are streaky, as shown in Fig. 1(a)-1(c) of the main manuscript, indicating a two-dimensional growth mode and zinc-blende structure maintained throughout the MBE growth.

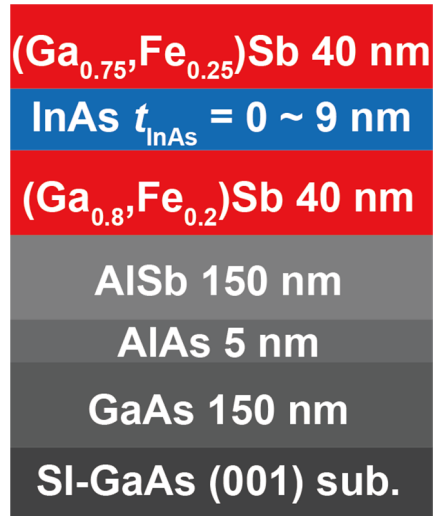

**Fig. S1.** Schematic sample structure examined in this study.

## 2. Magnetization of sample B

Magnetization curves of sample B measured at 10 K by superconducting quantum interference device (SQUID) magnetometry with a magnetic field applied perpendicular and parallel to the film plane are shown in Fig S2(a). Temperature dependence of the magnetization under an external magnetic field of 50 Oe is shown in Fig S2(b). These data indicate that the (Ga,Fe)Sb layers have perpendicular magnetic anisotropy at low temperature (10 K) as designed.[31] Fig. S2(c) shows an Arrot plot of sample B from the magnetic field dependence of magnetic circular dichroism (MCD – $H$ curves) [Fig. 4(b)], which reflects the magnetization of the top (Ga,Fe)Sb layer. $T_C$ estimated from the Arrot plot is around 260 K.

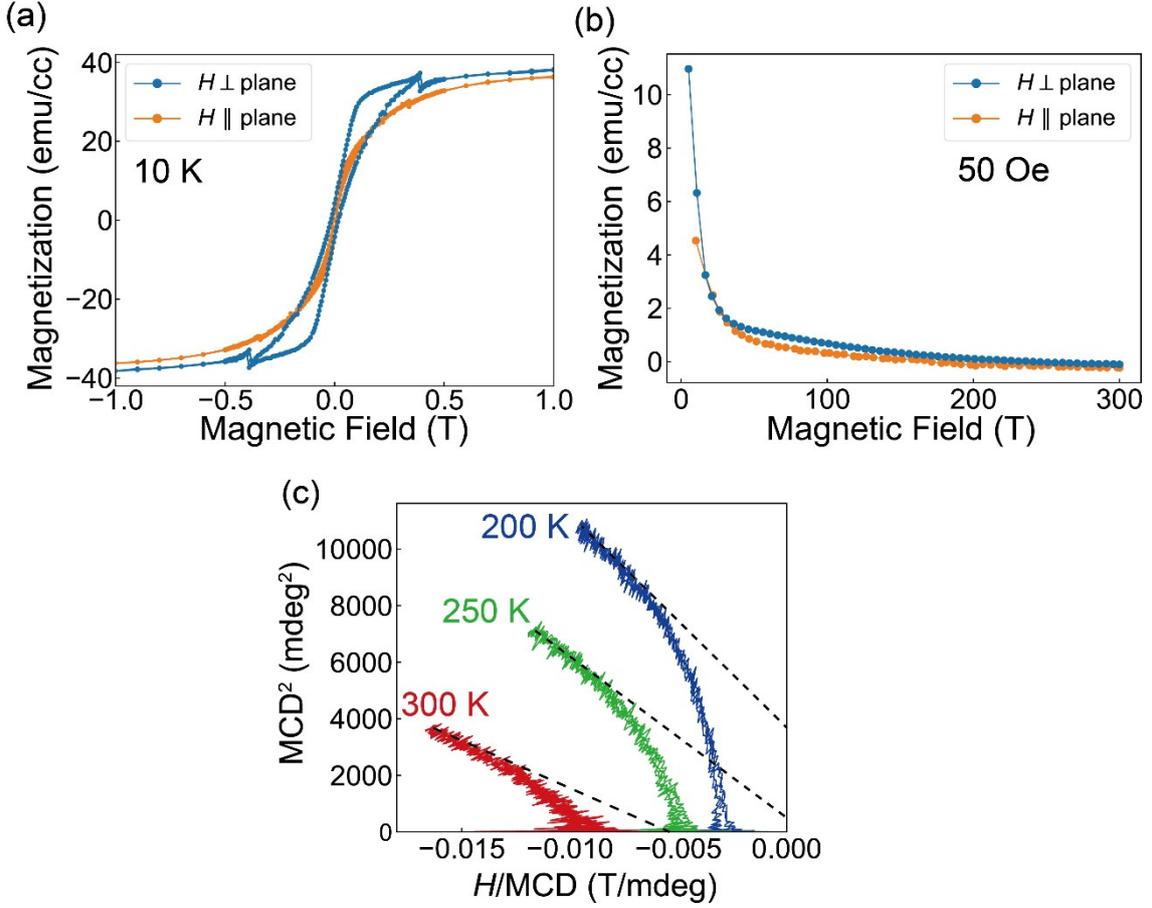

**Fig. S2.** (a) Magnetization hysteresis at 10 K and (b) temperature dependence of the magnetization (under an external magnetic field of 50 Oe) in sample B, measured by SQUID. Magnetic field is applied perpendicular (blue) and parallel (orange) to the film plane. (c) Arrot plot of sample B from the MCD – $H$ curves [Fig. 4(b)], which reflects the magnetization of the top (Ga,Fe)Sb layer. $T_C$ is estimated to be 260 K.

## 3. Removing the background MR from the raw MR curves

We removed the background MR from the raw MR curves [Fig. 3(a)] in order to estimate the spin-valve $MR_{SV}$ ratio, as shown in Fig. S3. We assumed the positive background MR is linear in sample A.[34] The linear background MR of sample A is removed so that the MR ratio is constant at high magnetic field (> 7 kOe). The negative background MR in sample C and D is fitted using the modified Khosla-Fischer model[32] as follows and subtracted from the raw MR curves.

$$\frac{R(H) - R(0)}{R(0)} = -a^2 \ln(1 + b^2 H^2) + \frac{c^2 H^2}{1 + d^2 H^2}$$

Here, $a$, $b$, $c$, and $d$ are fitting parameters. Resistances at high magnetic field (> 5 kOe)

were used for the fitting. Meanwhile, the MR curve of sample B has almost no background MR, possibly because the positive and negative background MRs cancel each other.

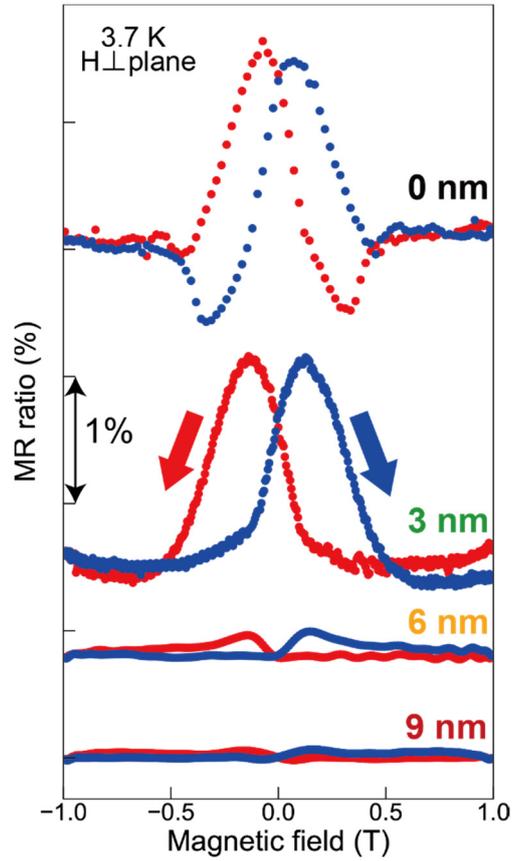

**Fig. S3.** MR curves at 3.7 K of the spin-valve structures with various $t_{\text{InAs}}$ (= 0, 3, 6, and 9 nm) after the background MR was removed from the raw MR curves shown in Fig. 3(a). The magnetic field is applied perpendicular to the film plane. The red and blue curves are the major loops with magnetic-field sweeping directions of + to – and – to +, respectively.